\documentclass[a4paper,12pt]{article}

\usepackage[utf8]{inputenc}

\usepackage{authblk}

\usepackage[margin=2.5cm]{geometry}

\usepackage{setspace}
\usepackage[shortlabels]{enumitem}

\usepackage{graphicx}
\graphicspath{ {./figures/} }

\usepackage[font=footnotesize,labelfont=bf,subrefformat=parens]{subcaption}

\usepackage{comment}

\usepackage{xcolor}

\usepackage{hyperref}
\hypersetup{
    colorlinks,
    linkcolor={red!50!black},
    citecolor={blue!50!black},
    urlcolor={blue!80!black}
}

\usepackage{amsmath}

\usepackage{amssymb}

\usepackage[lighttt]{lmodern}
\ttfamily
\DeclareFontShape{OT1}{lmtt}{m}{it}
     {<->sub*lmtt/m/sl}{}

\usepackage{listings}
\definecolor{codegreen}{rgb}{0,0.6,0}
\definecolor{codegray}{rgb}{0.5,0.5,0.5}
\definecolor{codepurple}{rgb}{0.58,0,0.82}
\definecolor{backcolour}{rgb}{0.95,0.95,0.92}
\lstdefinestyle{mystyle}{
    keywordstyle=\color{black}\bfseries, 
    numberstyle=\tiny\color{codegray},
    stringstyle=\color{black}\slshape,  
    basicstyle=\ttfamily\footnotesize,
    numbers=left,
    showspaces=false,
    showstringspaces=false,
}
\usepackage[british]{babel}
\usepackage{booktabs}

\title{Analytical solution of the Poiseuille flow of a De Kee viscoplastic fluid}

\author{Alexandros Syrakos}
\author{Aggelos Charalambous}
\affil{Department of Mechanical and Manufacturing Engineering, University of Cyprus, P.O. Box 20537,
1678 Nicosia, Cyprus}

\author{Georgios C. Georgiou}
\affil{Department of Mathematics and Statistics, University of Cyprus, PO Box 20537, 1678, Nicosia,
Cyprus}

\begin{document}

\newcommand{\vf}[1]{\underline{#1}}
\newcommand{\tf}[1]{\underline{\underline{#1}}}
\newcommand{\pd}[2]{\frac{\partial #1}{\partial #2}}
\newcommand{\ucd}[1]{\overset{\scriptscriptstyle \triangledown}{\tf{#1}}}
\newcommand{\gd}{\dot{\gamma}}
\newcommand\Rey{\operatorname{\mathit{Re}}}
\newcommand\Str{\operatorname{\mathit{Sr}}}
\newcommand\Wei{\operatorname{\mathit{Wi}}}
\newcommand\Deb{\operatorname{\mathit{De}}}
\newcommand\Bin{\operatorname{\mathit{Bn}}}

\maketitle

\begin{abstract}
We provide an explicit analytical solution of the planar Poiseuille flow of a viscoplastic fluid
governed by the constitutive equation proposed by De Kee and Turcotte (Chem.\ Eng.\ Commun.\ 6
(1980) 273--282). Formulae for the velocity and the flow rate are derived, making use of the Lambert
W function. It is shown that a solution does not always exist because the flow curve is bounded
from above and hence the rheological model can accommodate stresses only up to a certain limit. In
fact, the flow curve reaches a peak at a critical shear rate, beyond which it exhibits a negative
slope, giving rise to unstable solutions.
\end{abstract}

\section{Introduction}
\label{sec: introduction}

Viscoplastic fluids are characterised by the property of behaving in a solid-like manner when the
applied stress is below a limit value called the yield stress \cite{Coussot_2017, Balmforth_2014}.
Examples of such fluids include toothpaste, hair gel, mayonnaise, shaving foam, mud, mucus, clay,
fresh concrete, crude oil, and many others. This class of fluids includes a variety of materials
such as foams, emulsions, colloids, and physical gels, with the emergence of yield stress as a
macroscopic property being attributable to a variety of microscopic mechanisms, possibly different
for each material type \cite{Bonn_2017}.

Mathematical modelling of the rheological behaviour of viscoplastic fluids is a field that has been
developing during the last century or so. Classic viscoplastic models originate in the work of
Eugene Bingham who proposed the famous constitutive model that carries his name \cite{Bingham_1922}.
These models, commonly called \textit{simple yield stress fluids}, assume the material to have a
solid state that is completely rigid, and a fluid state which is that of a generalised Newtonian
fluid. The most popular such model, which incorporates both a yield stress and shear thinning or
thickening, is the Herschel-Bulkley (HB) model \cite{Herschel_1926}:

\begin{equation}
\label{eq: HB}
  \left\{
    \begin{array}{ll}
        \gd \;=\; 0  &
                                                                            \quad \tau < \tau_0 \\
        \tf{\tau} \;=\; \left( \dfrac{\tau_0}{\gd} \;+\;
                        k \gd^{n-1} \right) \tf{\gd} &
                                                                            \quad \tau \geq \tau_0
  \end{array}
  \right.
\end{equation}
where $\tf{\gd}$ is the rate-of-strain tensor and $\gd = (\tf{\gd} : \tf{\gd} \,/\, 2)^{1/2}$ is its
magnitude, $\tf{\tau}$ is the deviatoric stress tensor and $\tau = (\tf{\tau} : \tf{\tau} \,/\,
2)^{1/2}$ is its magnitude, $\tau_0$ is the yield stress, $k$ is the consistency index, and the
exponent $n$ determines if the fluid is shear-thinning ($n < 1$) or shear-thickening ($n > 1$). For
$n = 1$ the Herschel-Bulkley model reduces to the Bingham model. Another popular model of this class
is the Casson model \cite{Casson_1959}.

Real viscoplastic fluids exhibit additional rheological properties such as elasticity and thixotropy
\cite{Dinkgreve_2017, Larson_2019}, a fact that has given rise to recent efforts for the development
of more complicated rheological models with expanded physics \cite{Saramito_2009, Dimitriou_2019,
Varchanis_2019}. Nevertheless, simple yield stress fluids continue to be used at present and will
most likely persist in the future, having the advantages of simplicity and focus on plasticity and
shear-thinning, which are the defining aspects of many flows of interest. A recent defence of this
class of rheological models is provided in \cite{Frigaard_2019}.

Simple yield stress fluids are challenging from mathematical and computational perspectives. The
stress tensor is indeterminate in the unyielded (solid-state) regions, while the evolution of the
yield surfaces (the boundaries between the yielded and unyielded material) is not described
explicitly by some equation. Several numerical methods have been developed for solving the flows of
simple yield stress fluids \cite{Mitsoulis_2017, Saramito_2017, Moschopoulos_2022}, some of which
solve the original models directly while others first regularise them, effectively converting the
unyielded material into a very viscous fluid (something that may also have some physical
justification, at least for some materials). The most popular regularisation method is that of
Papanastasiou \cite{Papanastasiou_1987}.

A less popular simple yield stress fluid model was proposed by De Kee and Turcotte
\cite{DeKee_1980}:
\begin{equation}
\label{eq: DeKee}
  \left\{
    \begin{array}{ll}
        \gd \;=\; 0  &
                                                                            \quad \tau < \tau_0 \\
        \tf{\tau} \;=\; \left( \dfrac{\tau_0}{\gd} \;+\;
                        \eta_1 e^{-t_1 \gd} \right) \tf{\gd} &
                                                                            \quad \tau \geq \tau_0
  \end{array}
  \right.
\end{equation}
Compared to the Herschel-Bulkley model \eqref{eq: HB}, instead of the consistency $k$ and the
power-law exponent $n$, the De Kee model employs constants $\eta_1$ and $t_1$ which have units of
viscosity and time, respectively. For our analysis, it is convenient to define also the reciprocal
of the time constant as $\gd_1 = t_1^{-1}$, which has dimensions of strain rate, because, as will
be shown, this is a critical value of strain-rate that delimits distinct regions where the
properties of the model differ drastically. Like the Herschel-Bulkley model, the De Kee model can
predict both plasticity and shear-thinning.

In a later work \cite{Zhu_2005}, De Kee and co-workers presented a Papanastasiou-type regularised
version of the model in order to bound the viscosity at vanishing shear rate. Since the viscous
component $\eta_1 e^{-t_1 \gd}$ is already bounded -- which is an advantage of the De Kee model over
the HB model -- the regularisation needs to be applied only to the plastic component $\tau_0 / \gd$
(nevertheless, it should be pointed out that the HB viscosity is easily bounded by applying the
regularisation also to the viscous component $k\gd^{n-1}$ \cite{Sverdrup_2018, Syrakos_2020}).
Another advantage of the De Kee model is that the dimensions of its constants, $\eta_1$ and $t_1$,
are fixed, and they have a clear physical significance, in contrast to the HB parameters where the
dimensions of the consistency $k$ depend on $n$.

The De Kee -- Turcotte model has been used in various experimental and numerical studies.
Kaczmarczyk et al.\ \cite{Kaczmarczyk_2023} fitted the model \eqref{eq: DeKee}, incorporating
additional viscous terms $\eta_2 e^{-t_2 \gd}$ and $\eta_3 e^{-t_3 \gd}$, to rheological
measurements for \textit{Plantago ovata} water extract solutions. They examined both dilute (zero
yield stress) and semi-dilute (non-zero yield stress) solutions. Yahia and Khayat \cite{Yahia_2001}
made rheological measurements on cement grout and found that the De Kee model is suitable for
mixtures made of 100\% cement and rheology-modifying admixtures. Seo et al.\ \cite{Seo_2011}
proposed a generalised model for electrorheological fluids which reduces to the De Kee -- Turcotte
model for particular choices of parameters. The regularized version of the model \cite{Zhu_2005} was
used by Zare et al.\ \cite{Zare_2019} to model polymer blends and nanocomposites containing poly
(lactic acid), poly (ethylene oxide) and carbon nanotubes. In numerical studies, the model was used
for the numerical simulation of the cessation of viscoplastic Couette flow \cite{Zhu_2007} and the
numerical simulation of the flow in a rheometer with concentric cylinder geometry \cite{Wang_2011}.

The present work exposes an inherent limitation of the model: it only yields solutions within
limited parameter ranges. This limitation is due to its excessive shear thinning. As an application,
we will solve analytically the planar Poiseuille flow and determine the range of parameters for
which a solution exists. The solution is obtained with the use of the Lambert W function
\cite{Corless_1996}, which has proved quite useful in non-Newtonian fluid mechanics. This function
is briefly presented in Section \ref{sec: Lambert function}. The aforementioned limitation of the
model is exposed in Section \ref{sec: flow curves}, and the analytical solutions, both stable and
unstable, of planar Poiseuille flow are presented in Section \ref{sec: Poiseuille flow}.

\section{The Lambert W function}
\label{sec: Lambert function}

\begin{figure}[tb]
  \centering
  \includegraphics[width=12cm]{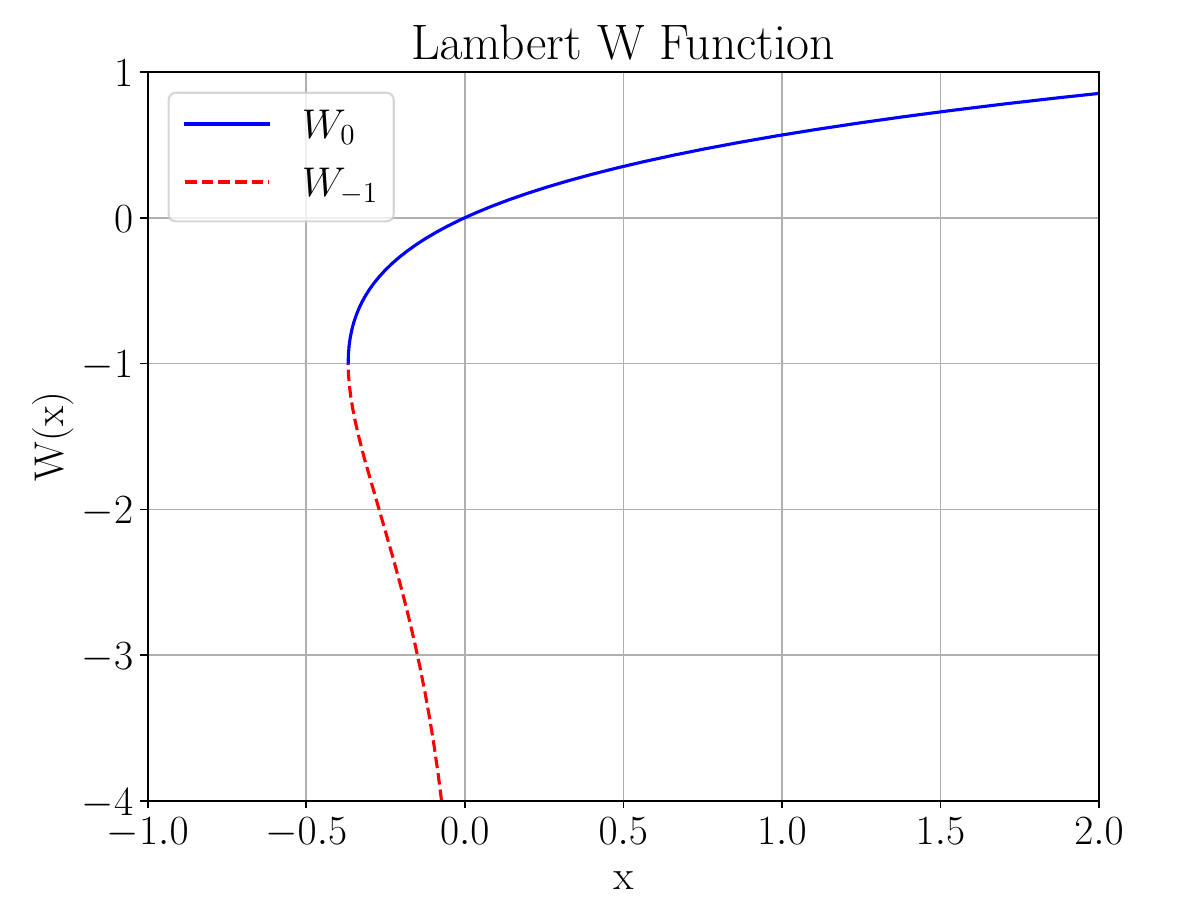}
  \caption{The Lambert W function, with its two branches, $W_0$ and $W_{-1}$.}
  \label{fig: Lambert function}
\end{figure}

Our analytical solution makes use of the Lambert W function, which returns the solution of the
equation $x e^x = y$:
\begin{equation}
\label{eq: Lambert}
  x \, e^x \;=\; y \;\;\Leftrightarrow\;\; x \;=\; W(y)
\end{equation}
The function is plotted in Fig.\ \ref{fig: Lambert function}. It is multivalued in the interval
$(-1/e, 0)$, and therefore consists of two branches: the principle branch, denoted by $W_0(x)$, for
$x \in [-1/e, \infty)$, and the secondary branch, denoted as $W_{-1}(x)$, for $x \in [-1/e,0)$.
These branches are illustrated in Fig.\ \ref{fig: Lambert function}; the principal branch is
strictly increasing from $-1$ to infinity, while the secondary branch is strictly decreasing from
$-1$ to minus infinity. It is important to note that the equation \eqref{eq: Lambert} does
\textit{not} have a solution for $y < -1/e$. Hence, $W(x)$ is not defined for $x < -1/e$.

An overview of the Lambert W function and its applications can be found in \cite{Corless_1996}. The
function is useful in Newtonian fluid mechanics \cite{Pitsillou_2020a}, and more so in non-Newtonian
fluid mechanics where it has many applications \cite{Pitsillou_2020b, Huilgol_2022}.

Some integrals involving the Lambert function that are useful for the present application are:

\begin{equation}
\label{eq: Lambert integral}
  \int W(x) \, dx
  \;=\;
  x \, W(x) \;-\; x \;+\; e^{W(x)} \;+\; c
\end{equation}

\begin{equation}
\label{eq: Lambert x integral}
  \int x \, W(x) \, dx
  \;=\;
  \frac{1}{8} \left( 2 W(x)^2 + 1 \right) \left( 2 W(x) - 1 \right) e^{2W(x)} \;+\; c
\end{equation}

\begin{equation}
\label{eq: Lambert exp integral}
  \int e^{W(x)} \, dx
  \;=\;
  \frac{1}{4} \left( 1 + 2W(x) \right) e^{2W(x)} \;+\; c
\end{equation}
where $c$ is an arbitrary constant of integration. All of these expressions can be obtained by
making the substitution $W(x) = v \Leftrightarrow v e^v = x \Rightarrow dx = e^v(1+v) dv$, and then
repeatedly performing integration by parts.

\section{Flow curve}
\label{sec: flow curves}

\begin{figure}[tb]
  \centering
  \includegraphics[width=12cm]{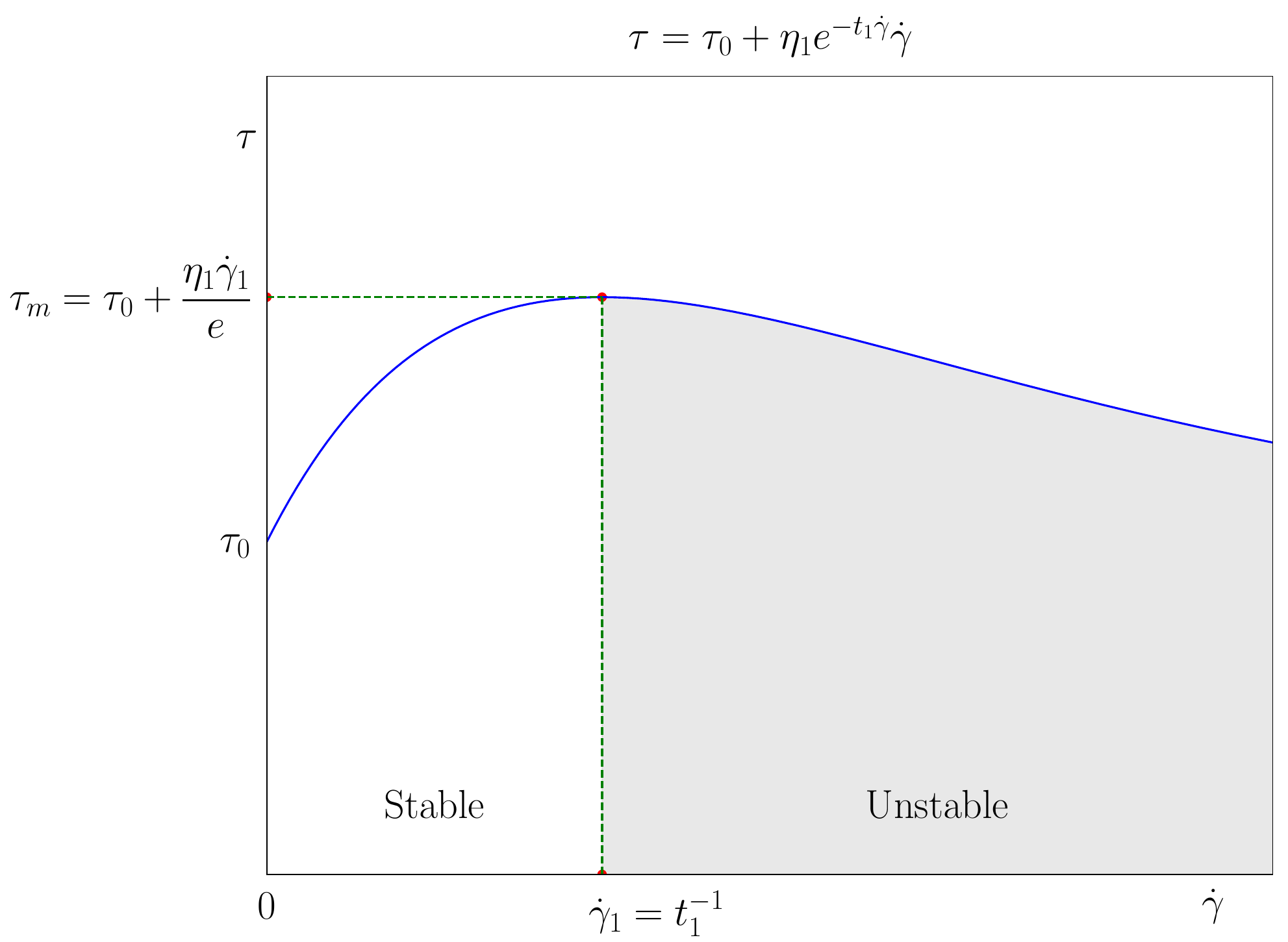}
  \caption{Variation of stress with the strain rate, for the De Kee model.}
  \label{fig: flow curve stress}
\end{figure}

\begin{figure}[tb]
  \centering
  \includegraphics[width=8.5cm]{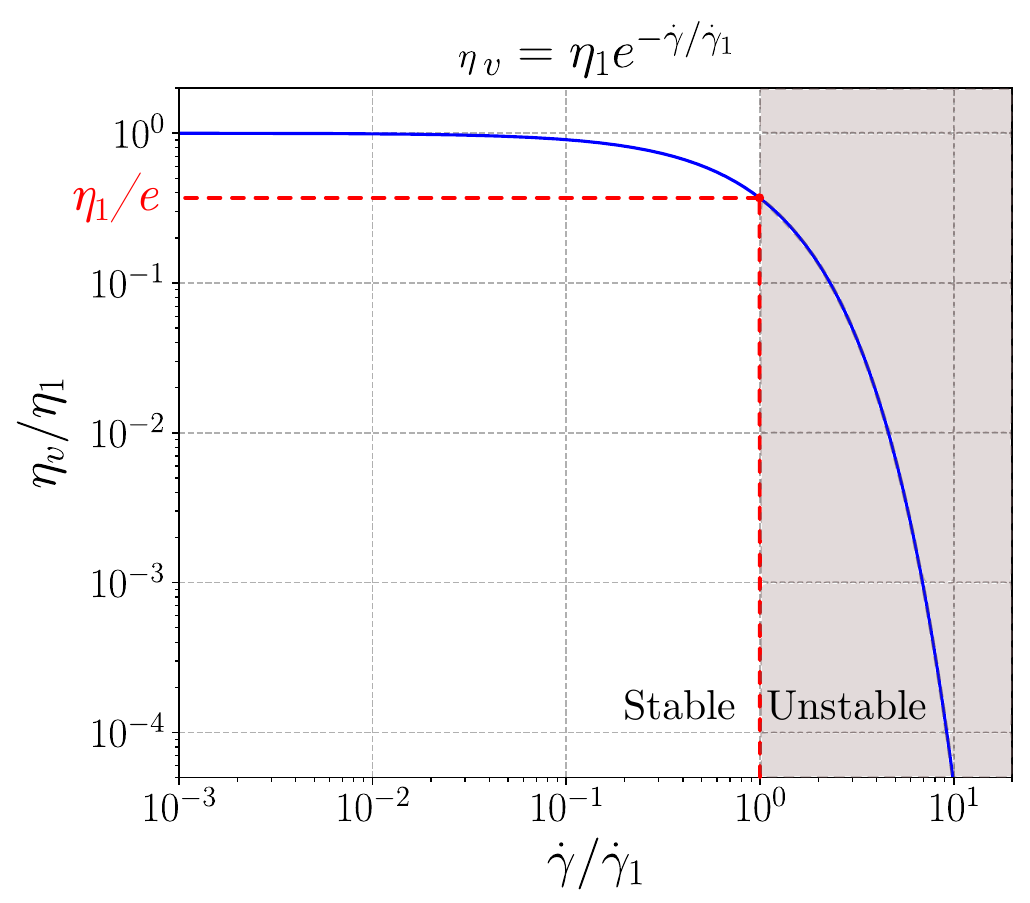}
  \caption{Variation of the ``viscous'' component of the viscosity, $\eta_1 e^{-t_1 \gd}$, with the
strain rate. The viscosity is normalised by $\eta_1$ and the strain rate by $\gd_1 = t_1^{-1}$.}
  \label{fig: flow curve viscosity}
\end{figure}

Figures \ref{fig: flow curve stress} and \ref{fig: flow curve viscosity} illustrate the variation
of stress and viscosity, respectively, with shear rate, for the De Kee -- Turcotte model. Plot
\ref{fig: flow curve stress} is obtained by taking the norm of Eq.\ \eqref{eq: DeKee}. In plot
\ref{fig: flow curve viscosity} only the ``viscous'' part, $\eta_1 e^{-t_1 \gd}$, of the viscosity
is considered -- the ``plastic'' part, $\tau_0 / \gd$, is omitted.

What is immediately striking in Fig.\ \ref{fig: flow curve stress} is that $\tau(\gd)$ is not a
strictly increasing function, but has a maximum of $\tau_m = \tau_0 + \eta_1\gd_1/e$ at $\gd =
\gd_1$. This is due to the excessive shear-thinning for $\gd > \gd_1$, which causes not only the
viscosity, but even the stress itself, to fall. This has the following repercussions.

Firstly, the stress cannot increase beyond the value $\tau_m$. This means that there are many
cases for which steady-state solutions do not exist because the momentum balance would require
higher stress values than the model can provide. One example is Poiseuille flows whose pressure
gradient exceeds a certain threshold, a case that will be examined shortly.

Secondly, for those cases that a solution exists, we can see that the same stress state can be
achieved with two different values of the shear rate -- that is, for each value of $\tau \in
(\tau_0, \tau_m)$ the corresponding horizontal line in Fig.\ \ref{fig: flow curve stress} intersects
the $\tau(\dot{\gamma})$ curve at two points, say $\dot{\gamma}_{-} < \dot{\gamma}_0$ and
$\dot{\gamma}_{+} > \dot{\gamma}_0$. Hence, we expect multiple solutions, when they exist. Of these
solutions, those with shear rates $\dot{\gamma} > \dot{\gamma}_0$ will be unstable
\cite{Yerushalmi_1970} because any perturbation in $\dot{\gamma}$ will cause the stress to change in
such a direction that will amplify further the change in the shear rate, starting a vicious circle.
If $\dot{\gamma}$ is increased, $\tau$ falls and the reduced viscous resistance in the fluid will
cause a further increase in $\dot{\gamma}$ and so on. The opposite will happen if $\dot{\gamma}$ is
perturbed in the negative direction: this will cause $\tau$ to increase, strengthening the viscous
resistance to the flow and causing further decrease to $\dot{\gamma}$ etc. These features of the
model will be demonstrated in the case of planar Poiseuille flow in the next section.

Concerning the viscosity, Fig.\ \ref{fig: flow curve viscosity} seems like a typical shear-thinning
viscosity curve. At low shear rates, $\gd \ll \gd_1$, the ``viscous'' contribution to the viscosity,
 $\eta_1 e^{-\gd / \gd_1}$, is approximately constant and equal to $\eta_1$, with the model
exhibiting an almost Newtonian behaviour, in contrast to the $k\gd^{n-1}$ component of the HB
viscosity which becomes infinite at vanishing shear rate. At $\gd = \gd_1$ this viscosity component
has decreased to $\eta_v = \eta_1 / e \approx 0.37\eta_1$. Beyond $\gd = \gd_1$ the shear thinning
intensifies dramatically. However, it should be noted that $\gd > \gd_1$ lies in the unstable regime
and therefore the lowest practically achievable viscosity is $\eta_v = \eta_1 / e \approx
0.37\eta_1$ with viscosities lower than that being practically impossible to achieve. In other
words, practically $\eta_v \in [0.37\eta_1, \eta_1]$.

To give a feel of the practical range of applicability of the model, Table \ref{table: rheo data}
lists the values of its parameters as fitted to rheological data for various real fluids by De Kee
and Turcotte \cite{DeKee_1980}, together with the corresponding values of critical rate of strain
$\gd_1$ and maximum attainable stress $\tau_m$.
\begin{table}
\begin{center}
\begin{tabular}[c]{l|lll|ll}\toprule
 & $\tau_0$ [Pa] & $\eta_1$ [Pa.s] & $t_1$ [s] & $\gd_1$ [1/s] & $\tau_m$ [Pa]\\\midrule
Banana puree & $1.04 \times 10^2$ & $6.26 \times 10^4$ & $6.23 \times 10^1$ & $1.61 \times 10^{-2}$
& $4.74 \times 10^2$\\
Blood & $3.81 \times 10^{-3}$ & $7.17 \times 10^{-3}$ & $3.29 \times 10^{-2}$ & $3.04 \times 10^1$ &
$8.40 \times 10^{-2}$\\
Mayonnaise & $1.35 \times 10^2$ & $4.20 \times 10^{-1}$ & $1.44 \times 10^{-4}$ & $6.94 \times 10^3$
& $1.21 \times 10^3$\\
Yogurt & $4.17 \times 10^1$ & $1.15 \times 10^{-2}$ & $4.52 \times 10^{-5}$ & $2.21 \times 10^4$ &
$1.35 \times 10^2$\\\bottomrule
\end{tabular}
\caption{Parameters of the model \eqref{eq: DeKee} as fitted to rheological data for various real
fluid by De Kee and Turcotte \cite{DeKee_1980}. The last two columns list the critical rate of
strain and the maximum attainable stress, respectively (Fig.\ \ref{fig: flow curve stress}).}
\label{table: rheo data}
\end{center}
\end{table}

\section{Planar Poiseuille flow}
\label{sec: Poiseuille flow}

With these considerations, let us proceed to the analytical solution of planar Poiseuille flow.

\subsection{Preliminary considerations}
\label{ssec: preliminary}

For simple shear flow, the De Kee model \eqref{eq: DeKee} reduces to the following form:

\begin{equation}
\label{eq: DeKee shear}
  \left\{
  \begin{array}{ll}
    \dfrac{du}{dy} \;=\; 0  &
                                                                      \quad |\tau_{yx}| < \tau_0 \\
    \tau_{yx} \;=\; \left( \dfrac{\tau_0}{|du/dy|} \;+\;
                    \eta_1 e^{-\frac{|du/dy|}{\gd_1}} \right) \dfrac{du}{dy} &
                                                                      \quad |\tau_{yx}| \geq \tau_0
  \end{array}
  \right.
\end{equation}
where $x$ is the flow direction and $u$ is the velocity in that direction, $y$ is the perpendicular
direction across which $u(y)$ varies, and $\tau_{yx}$ is the shear stress.

\begin{figure}[tb]
  \centering
  \includegraphics[width=14cm]{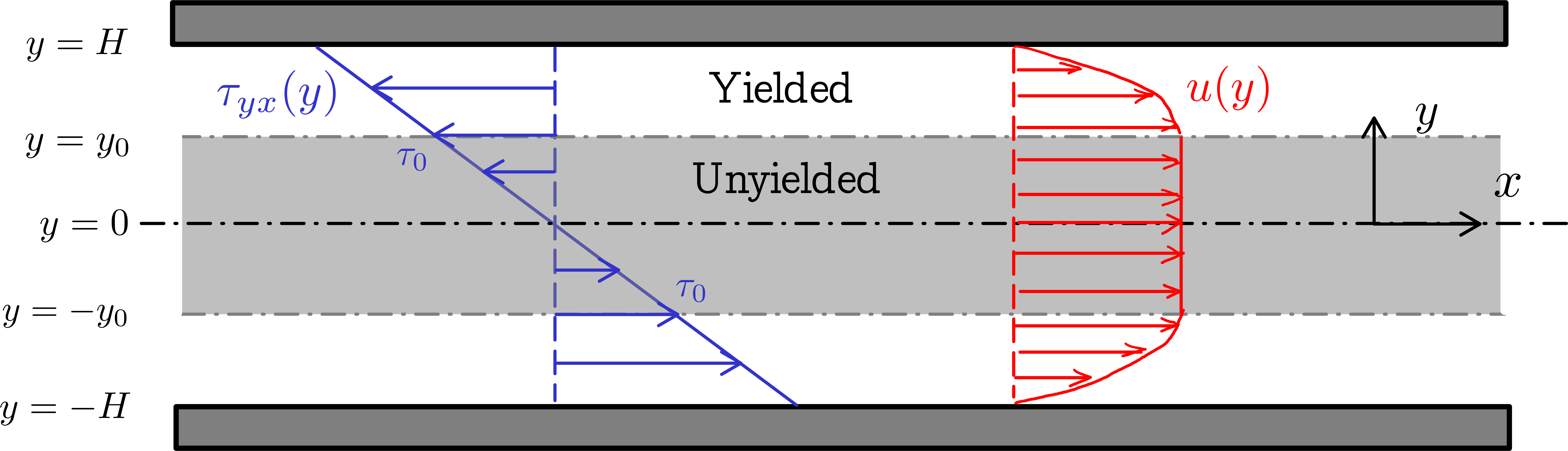}
  \caption{Sketch of the configuration of viscoplastic planar Poiseuille flow.}
  \label{fig: PF arrangement}
\end{figure}

One such flow is planar Poiseuille flow, a steady flow where fluid is pushed along a channel formed
by two infinite horizontal parallel plates, located at a distance $2H$ apart, by an imposed pressure
gradient, $G = -dp/dx$ (Fig. \ref{fig: PF arrangement}). For this flow, the momentum balance (Cauchy
equation) reduces to:
\begin{equation}
\label{eq: Cauchy}
  \tau_{yx} \;=\; -Gy
\end{equation}
where $y=0$ is set midway between the plates (Fig. \ref{fig: PF arrangement}). The stress grows
linearly with distance from the midplane, ranging from zero at the midplane to a maximum magnitude
of $GH$ at the plates. Therefore, as long as $\tau_0 \neq 0$, an unyielded core will form in the
middle of the domain, up to a distance of
\begin{equation}
\label{eq: y0}
  y_0 \;=\; \frac{\tau_0}{G}
\end{equation}
from the midplane (from Eq.\ \eqref{eq: Cauchy}). If $y_0 > H$ then obviously no flow will occur
and the whole material will be unyielded, provided that no-slip conditions apply at the plates,
which is an assumption that will be made in the present paper. Otherwise, if $y_0 < H$, then there
will be a yielded zone in $y \in [y_0, H]$ and flow will occur. We can therefore focus on the
partially yielded case and first consider the yielded zone, where we can substitute Eq.\ \eqref{eq:
DeKee shear} in Eq.\ \eqref{eq: Cauchy}; in the upper half ($y>0$) of the domain, where $du/dy < 0$,
this substitution gives
\begin{equation}
\label{eq: DE}
  \left(
  \frac{\tau_0}{-du/dy}
  \;+\;
  \eta_1 \, e^{\frac{du/dy}{\gd_1}}
  \right) \frac{du}{dy}
  \;=\;
  -Gy
\end{equation}
This can be easily manipulated into the form
\begin{equation}
\label{eq: DE manipulated}
  \frac{1}{\gd_1} \frac{du}{dy} \, e^{\frac{1}{\gd_1} \frac{du}{dy}}
  \;=\;
  \frac{\tau_0 - Gy}{\eta_1 \,\gd_1}
\end{equation}
Applying the Lambert function to both sides and rearranging we get
\begin{equation}
\label{eq: dudy}
  \frac{du}{dy} \;=\; \gd_1 \, W\left( \frac{\tau_0 - Gy}{\eta_1 \, \gd_1} \right)
\end{equation}

The left-hand side, $du/dy$, is negative, and hence the output of the Lambert function on the
right-hand side must also be negative, which requires that its argument is negative (Fig.\
\ref{fig: Lambert function}). This is indeed the case, as the occurrence of flow means that $Gy >
\tau_0$ (Eq.\ \eqref{eq: Cauchy}). But this means that we are in the region where $W$ has two
branches, and hence there are two solutions to Eq.\ \eqref{eq: dudy}, one employing $W_0$ and one
employing $W_{-1}$. We will examine this issue later, but for now let us proceed without
particularising the branch that is selected.

According to what was said in Sec.\ \ref{sec: Lambert function}, in order for Eq.\ \eqref{eq: dudy}
to have a solution, the argument of the Lambert function must be greater than or equal to $-1/e$.
This argument is negative, and its magnitude is maximised at $y = H$. Therefore, the existence of a
solution requires that
\begin{equation}
\label{eq: condition 1}
  -\frac{1}{e} \;\leq\; \frac{\tau_0 - GH}{\eta_1 \, \gd_1}
  \quad\Leftrightarrow\quad
  GH \;\leq\; \tau_0 \;+\; \frac{\eta_1 \gd_1}{e}
\end{equation}
Now, from Eq.\ \eqref{eq: Cauchy}, $GH$ is the maximum stress value, required at the plates so
that the pressure gradient is counterbalanced and the flow is steady. Therefore, Eq.\ \eqref{eq:
condition 1} is equivalent to the condition
\begin{equation}
\label{eq: condition 2}
  \tau_{xy}(y=H) \;\leq\; \tau_0 \;+\; \frac{\eta_1 \gd_1}{e} \;\equiv\; \tau_m
\end{equation}
which simply says that the stress should be everywhere smaller than the maximum value $\tau_m$
producible by the De Kee -- Turcotte model, as shown in Sec.\ \ref{sec: flow curves}. If the
pressure gradient $G$ is too large for $\tau_m$ to counteract it (i.e.\ condition \eqref{eq:
condition 1} is not satisfied), then steady-state flow is not possible.

\subsection{Velocity profile}
\label{ssec: velocity}

Assuming that the pressure gradient is sufficiently small to satisfy condition \eqref{eq: condition
1}, we will proceed with the integration of Eq.\ \eqref{eq: dudy}. For convenience, it will be
brought to non-dimensional form by employing the following non-dimensionalisation:
\begin{equation}
\label{eq: non dimensionalisation}
  \tilde{u} = \frac{u}{\gd_1 H}
  \quad , \quad
  \tilde{y} = \frac{y}{H} \;\Rightarrow\; \frac{d}{dy} = \frac{1}{H}\frac{d}{d\tilde{y}}
  \quad , \quad
  \tilde{G} = \frac{G H}{\eta_1 \gd_1}
\end{equation}
We will also substitute $\tau_0 = G y_0$ (Eq.\ \eqref{eq: y0}). The non-dimensional form of
\eqref{eq: dudy} is then:
\begin{equation}
\label{eq: dudy ND}
  \frac{d \tilde{u}}{d \tilde{y}}
  \;=\;
  W \left( \tilde{G}(\tilde{y}_0 - \tilde{y}) \right)
\end{equation}
This can be integrated using Eq.\ \eqref{eq: Lambert integral}, while the constant of integration
can be determined by the boundary condition $\tilde{u}(H) = 0$, to arrive at the following velocity
profile for the yielded region:

\begin{equation}
\label{eq: velocity yielded}
\begin{aligned}
  \tilde{u}
  \;&=\;
  \left( \tilde{y} - \tilde{y}_0 \right)
    \left[ W\left( \tilde{G} \left( \tilde{y}_0 - \tilde{y} \right)\right) - 1 \right]
  \;-\;
  \left( 1 - \tilde{y}_0 \right)
    \left[ W\left( \tilde{G} \left( \tilde{y}_0 - 1 \right)\right) - 1 \right]\\[0.25cm]
  \;&+\;
  \frac{1}{\tilde{G}} \left[
    e^{W\left( \tilde{G} \left( \tilde{y}_0 - 1 \right)\right)}
    \;-\;
    e^{W\left( \tilde{G} \left( \tilde{y}_0 - \tilde{y} \right)\right)}
    \right]
  \;,\qquad \tilde{y} \in [\tilde{y}_0,1]
\end{aligned}
\end{equation}
In the unyielded region the velocity is uniform and equal to that of the yielded region at
$\tilde{y} = \tilde{y}_0$:
\begin{equation}
\label{eq: velocity unyielded}
  \tilde{u}
  \;=\;
  \frac{1}{\tilde{G}} \left[ e^{W\left( \tilde{G} \left( \tilde{y}_0 - 1 \right)\right)} - e^{W(0)}
\right]
  \;-\;
  \left( 1 - \tilde{y}_0 \right)
    \left[ W\left( \tilde{G} \left( \tilde{y}_0 - 1 \right)\right) - 1 \right]
  \;, \qquad \tilde{y} \in [0,\tilde{y}_0]
\end{equation}

At this point, it is pertinent to consider the issue of the branches of $W$. Returning to Eq.\
\eqref{eq: dudy}, we note the following.

\subsubsection*{Principal branch}

For $x \leq 0$, $W_0(x) \in [-1,0]$, so that it follows from Eq.\ \eqref{eq: dudy} that $|du/dy|
\leq \gd_1$ and we are in the stable region of Fig.\ \ref{fig: flow curve stress}.

With increasing $y$, i.e.\ closer to the wall, the argument $(\tau_0 - Gy)/ \eta_1 \gd_1$ of $W$ in
Eq.\ \eqref{eq: dudy} becomes more negative, and hence $|W_0(\cdot)|$ increases (Fig.\ \ref{fig:
Lambert function}), implying that $|du/dy|$ also increases (Eq.\ \eqref{eq: dudy}). This is normal
behaviour: higher velocity gradients develop near the walls; it is because higher stresses require
higher velocity gradients in the stable region.

On the other hand, considering what happens when $y \rightarrow y_0$, due to Eq.\ \eqref{eq: y0}
the argument of $W_0$ in Eq.\ \eqref{eq: dudy} tends to zero, and so does $W_0(\cdot)$ itself, so
that the velocity gradient is zero at the yield surface -- again, normal behaviour exhibited by
other viscoplastic models as well. Stress continuity at the yield surface requires that the viscous
part of the stress reduces towards zero, leaving only the plastic part, as we approach the yield
surface from the yielded side.

\begin{figure}[tb]
    \centering
    \begin{subfigure}[b]{0.49\textwidth}
        \centering
        \includegraphics[height=6.5cm]{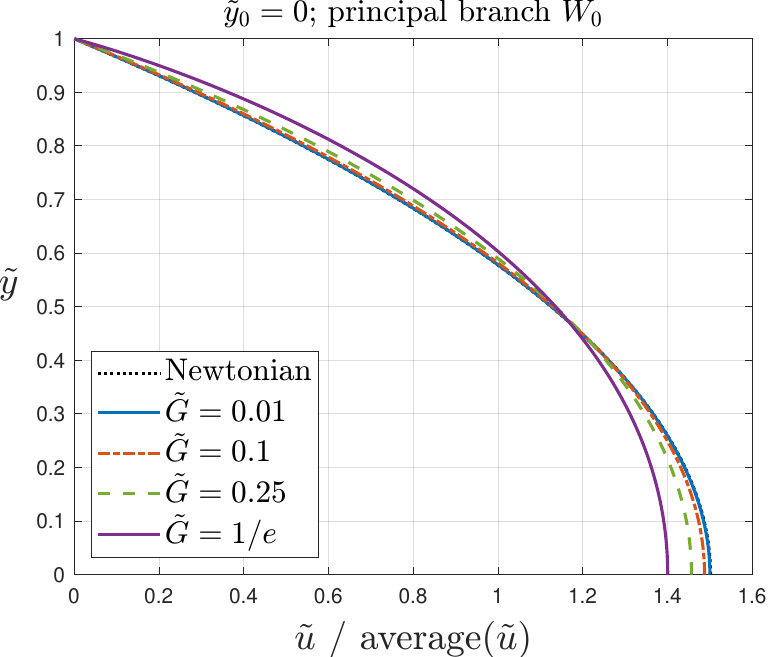}
        \caption{}
        \label{sfig: velocity profile y0=0 W0}
    \end{subfigure}
    \begin{subfigure}[b]{0.49\textwidth}
        \centering
        \includegraphics[height=6.5cm]{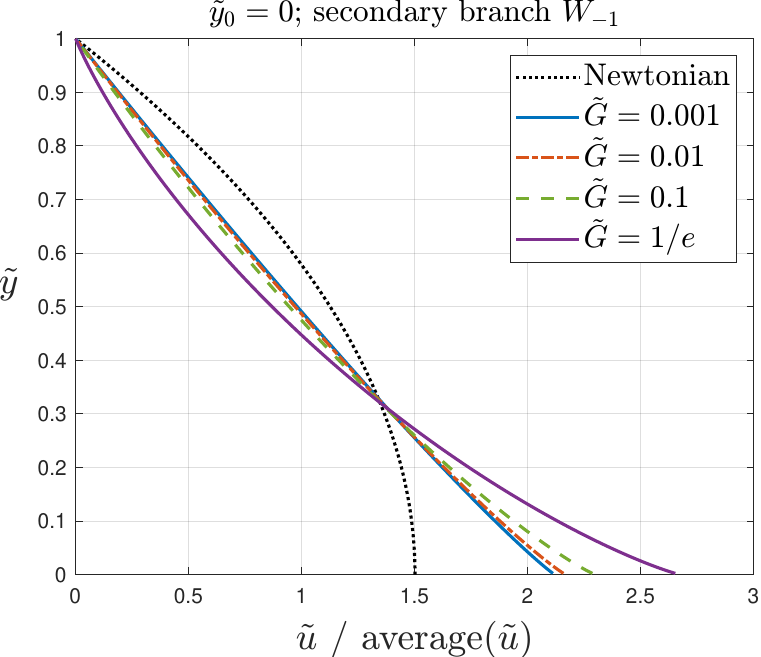}
        \caption{}
        \label{sfig: velocity profile y0=0 W-1}
    \end{subfigure}
    \caption{Velocity profiles, normalised by the mean velocity, for $y_0 = 0$ (no yield stress),
obtained with \subref{sfig: velocity profile y0=0 W0} $W_0$ (stable) and \subref{sfig: velocity
profile y0=0 W-1} $W_{-1}$ (unstable).}
  \label{fig: velocity profiles y0=0}
\end{figure}

Velocity profiles for cases without yield stress ($\tilde{y}_0 = 0$) and with yield stress
($\tilde{y}_0 = 0.5$) are shown in Figs.\ \ref{sfig: velocity profile y0=0 W0} and \ref{sfig:
velocity profile y0=0.5 W0}, respectively, for various values of dimensionless pressure gradient
$\tilde{G}$, up to the maximum allowable for steady-state attainment ($\tilde{G}_{\mathrm{max}} =
1/e$ for $\tilde{r}_0 = 0$ and $\tilde{G}_{\mathrm{max}} = 2/e$ for $\tilde{r}_0 = 0.5$). The
maximum allowable dimensionless pressure gradient is obtained by substituting $\tau_0 = G r_0$ in
the condition \eqref{eq: condition 1} and non-dimensionalising it to get:

\begin{equation}
\label{eq: condition ND}
  \tilde{G} \left( 1 - \tilde{y}_0 \right) \;\leq\; \frac{1}{e}
\end{equation}

\begin{figure}[tb]
    \centering
    \begin{subfigure}[b]{0.49\textwidth}
        \centering
        \includegraphics[height=6.5cm]{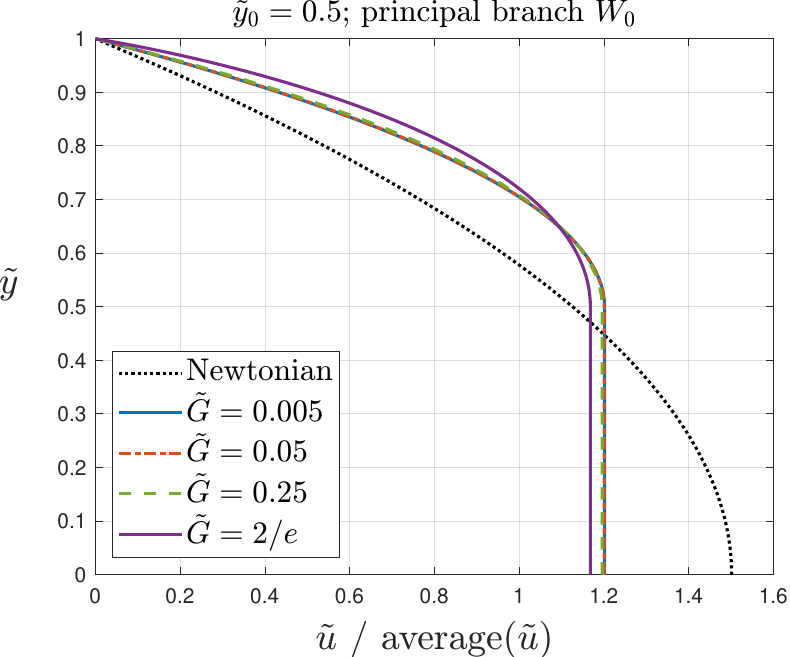}
        \caption{}
        \label{sfig: velocity profile y0=0.5 W0}
    \end{subfigure}
    \begin{subfigure}[b]{0.49\textwidth}
        \centering
        \includegraphics[height=6.5cm]{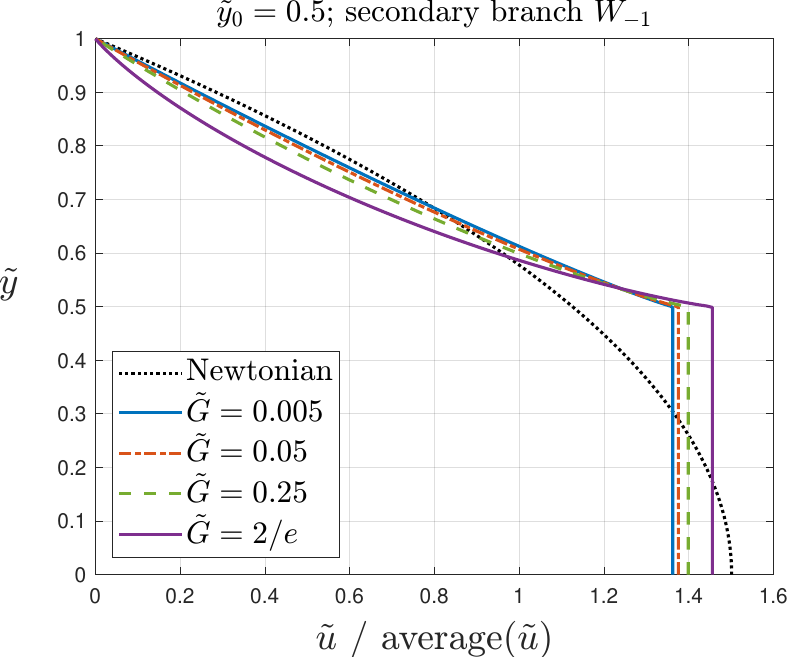}
        \caption{}
        \label{sfig: velocity profile y0=0.5 W-1}
    \end{subfigure}
    \caption{Velocity profiles, normalised by the mean velocity, for $y_0 = 0.5$ obtained with
\subref{sfig: velocity profile y0=0 W0} $W_0$ (stable) and \subref{sfig: velocity
profile y0=0 W-1} $W_{-1}$ (unstable).}
  \label{fig: velocity profiles y0=0.5}
\end{figure}

\subsubsection*{Secondary branch}

Since $W_{-1}(\cdot) \leq -1$, it follows from Eq.\ \eqref{eq: dudy} that $|du/dy| \geq \gd_1$ and
we are in the unstable region of Fig.\ \ref{fig: flow curve stress}.

With increasing $y$, i.e.\ closer to the wall, the argument $(\tau_0 - Gy)/ \eta_1 \gd_1$ of $W$ in
Eq.\ \eqref{eq: dudy} becomes more negative, and hence $|W_{-1}(\cdot)|$ \textit{decreases} (Fig.\
\ref{fig: Lambert function}), implying that $|du/dy|$ \textit{decreases as well} (Eq.\ \eqref{eq:
dudy}). This is the opposite of what is normally expected, but due to $\tau_{yx}(\gd)$ being a
decreasing function in the unstable region (Fig.\ \ref{fig: flow curve stress}), in order to get the
needed higher stresses near the wall the shear rate has to decrease there.

Again counterintuitively, when $y \rightarrow y_0$, due to Eq.\ \eqref{eq: y0} the argument of
$W_{-1}$ in Eq.\ \eqref{eq: dudy} tends to zero, and $W_{-1}(\cdot)$ tends to $-\infty$, so
that the velocity gradient becomes infinite at the yield surface.

Velocity profiles for cases without yield stress ($\tilde{y}_0 = 0$) and with yield stress
($\tilde{y}_0 = 0.5$) are shown in Figs.\ \ref{sfig: velocity profile y0=0 W-1} and \ref{sfig:
velocity profile y0=0.5 W-1}, respectively, for various values of dimensionless pressure gradient
$\tilde{G}$, up to the maximum allowable for steady-state attainment. Of course, these solutions
are unstable. The plots seem to show finite values of the velocity gradient \eqref{eq: dudy ND} at
$\tilde{y} \rightarrow \tilde{y}_0$ instead of the theoretical infinite one, but this is due to the
slowness of the decrease of $W_{-1}(x)$ towards $-\infty$ as $x \rightarrow 0$. For example, for
$\tilde{y}_0 - \tilde{y} = 0.001$, which is a typical $\tilde{y}$-value resolution for drawing the
plots, Eq.\ \eqref{eq: dudy ND} gives a dimensionless velocity gradient $d\tilde{u} / d\tilde{y}$ of
$-14.2$ for $\tilde{G} = 0.01$ and $-11.7$ for $\tilde{G} = 0.1$ (note also that the slopes in
Figs.\ \ref{sfig: velocity profile y0=0 W-1} and \ref{sfig: velocity profile y0=0.5 W-1} are not
to scale, because the velocities have been normalized by their average values).

\subsection{Flow rate}
\label{ssec: flow rate}

\begin{figure}[tb]
    \centering
    \begin{subfigure}[b]{0.49\textwidth}
        \centering
        \includegraphics[width=0.99\linewidth]{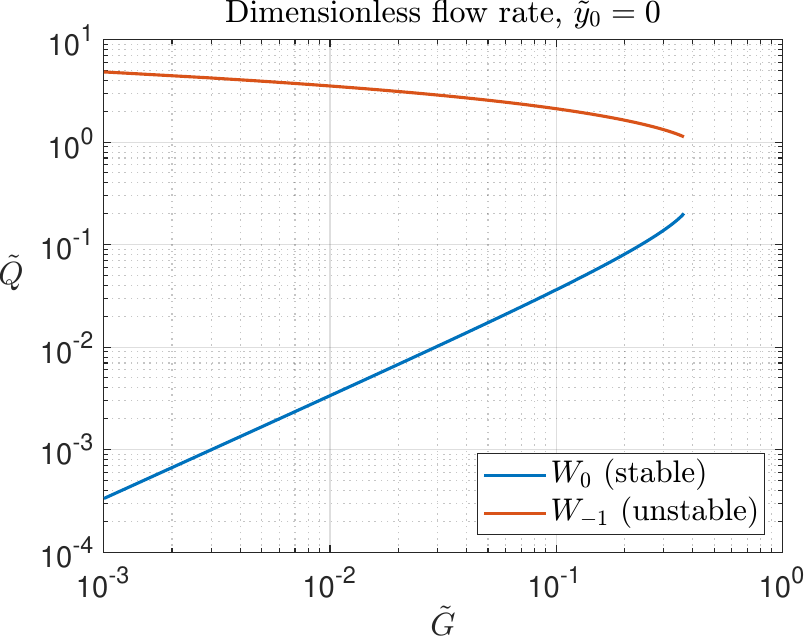}
        \caption{}
        \label{sfig: Q y0=0}
    \end{subfigure}
    \begin{subfigure}[b]{0.49\textwidth}
        \centering
        \includegraphics[width=0.99\linewidth]{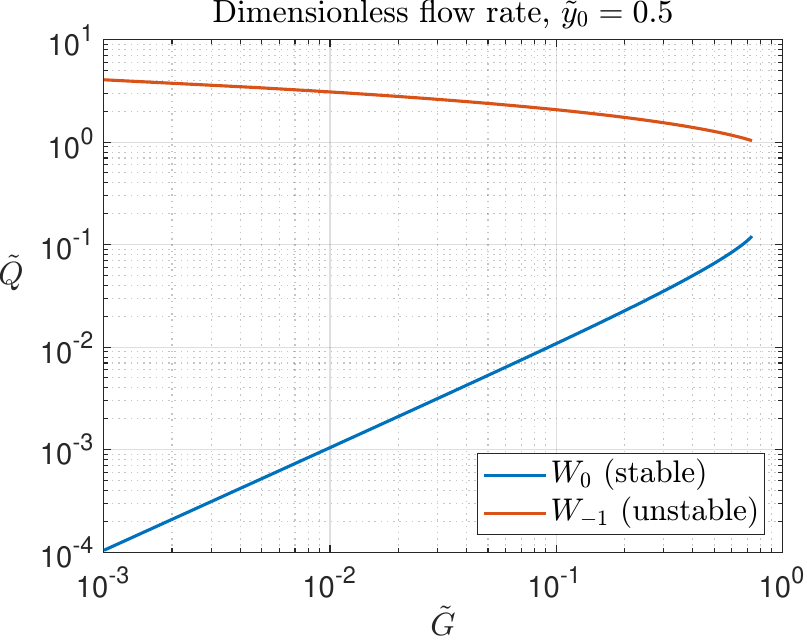}
        \caption{}
        \label{sfig: Q y0=0.5}
    \end{subfigure}
    \caption{Plots of dimensionless flow rate $\tilde{Q}$ as a function of the dimensionless
pressure gradient $\tilde{G}$, for the case that the fluid has no yield stress \subref{sfig: Q
y0=0} and for the viscoplastic case with $\tilde{y}_0$ held at 0.5 \subref{sfig: Q y0=0.5}.}
  \label{fig: Q}
\end{figure}

When a solution exists, the flow rate, per unit width, can be calculated by integrating the
velocity across the height of the channel:
\begin{equation}
\label{eq: Q dimensional}
  Q \;=\; \int_{y=-H}^{y=+H} u \, dy \;=\; 2 \int_{y=0}^{y=H} u \, dy
\end{equation}
where the symmetry about the $y=0$ plane has been exploited. Also, it will be convenient to
non-dimensionalise the flow rate:

\begin{equation}
\label{eq: Q step 1}
  \tilde{Q} \;\equiv\; \frac{Q}{2 \gd_1 H^2}
  \;=\;
  \int_{\tilde{y}=0}^{\tilde{y}=\tilde{y}_0} \tilde{u}_0 \, d\tilde{y}
  \;+\;
  \int_{\tilde{y}=\tilde{y}_0}^{\tilde{y}=1} \tilde{u} \, d\tilde{y}
\end{equation}
where, for convenience, the steady velocity \eqref{eq: velocity unyielded} of the unyielded plug is
denoted as $\tilde{u}_0$. With this notation, the velocity in the yielded region can be written as
\begin{equation}
\label{eq: velocity yielded 2}
  \tilde{u}
  \;=\;
  \tilde{u}_0
  \;+\;
  \left( \tilde{y} - \tilde{y}_0 \right)
    \left[ W\left( \tilde{G} \left( \tilde{y}_0 - \tilde{y} \right)\right) - 1 \right]
  \;-\;
  \frac{1}{\tilde{G}} \, e^{W\left( \tilde{G} \left( \tilde{y}_0 - \tilde{y} \right)\right)}
  \;+\;
  \frac{1}{\tilde{G}} e^{W(0)}
  \;, \quad \tilde{y} \in [\tilde{y}_0,1]
\end{equation}
and the above integral becomes
\begin{equation}
\label{eq: Q step 2}
\begin{aligned}
  \tilde{Q} \;&=\; \tilde{u}_0 \;+\; \frac{1-\tilde{y}_0}{\tilde{G}} \, e^{W(0)}
  \;+\;
  \int_{\tilde{y}=\tilde{y}_0}^{\tilde{y}=1}
    \left( \tilde{y} - \tilde{y}_0 \right)
    \left[ W\left( \tilde{G} \left( \tilde{y}_0 - \tilde{y} \right)\right) - 1 \right] d\tilde{y}
  \\[0.4cm]
  \;&-\;
  \frac{1}{\tilde{G}} \int_{\tilde{y}=\tilde{y}_0}^{\tilde{y}=1}
  e^{W\left( \tilde{G} \left( \tilde{y}_0 - \tilde{y} \right)\right)} d\tilde{y}
\end{aligned}
\end{equation}
This can be evaluated with the help of Eqs.\ \eqref{eq: Lambert x integral} and \eqref{eq: Lambert
exp integral}, to obtain
\begin{equation}
\label{eq: Q final}
\begin{aligned}
  \tilde{Q} \;&=\; \tilde{u}_0 \;+\; \frac{1-\tilde{y}_0}{\tilde{G}} \, e^{W(0)}
  \;+\;
  \frac{1}{8\tilde{G}^2} \left[
    \left( 2 \varpi^2 + 1 \right) \left( 2 \varpi - 1\right) e^{2 \varpi}
    \;+\; e^{W(0)} \right]
    \;-\; \frac{(1 - \tilde{y}_0)^2}{2}\\[0.2cm]
  \;&+\;
  \frac{1}{4\tilde{G}^2} \left[
    \left( 2 \varpi + 1 \right) e^{2\varpi} \;-\; e^{W(0)} \right]
\end{aligned}
\end{equation}
where $\varpi = W(\tilde{G}(\tilde{y}_0-1))$. Because $W_0(0) = 0$ and $W_{-1}(0) = -\infty$, the
term $e^{W(0)}$ equals $1$ for the stable branch and $0$ for the unstable one. Figure \ref{fig: Q}
shows plots of the dimensionless flow rate $\tilde{Q}$ as a function of the dimensionless pressure
gradient $\tilde{G}$ for $\tilde{y}_0 = 0$ (no yield stress) and $\tilde{y}_0 = 0.5$. Note that in
the latter case, since $\tilde{y}_0 = 0.5$ is held constant in Fig.\ \ref{sfig: Q y0=0.5}
irrespective of $\tilde{G}$, the curve $\tilde{Q} = f(\tilde{G})$ should not be construed as
varying the pressure gradient in a fixed channel with a fixed fluid, but in order for $\tilde{y}_0$
to remain constant as the pressure gradient varies the yield stress of the fluid must vary
simultaneously with $\tilde{G}$ (from Eq.\ \eqref{eq: y0} we get $\tilde{y}_0 = \tilde{\tau}_0 /
\tilde{G}$ where $\tilde{\tau}_0 = \tau_0 / \eta_1 \dot{\gamma}_1$).

\section{Conclusions}
\label{sec: conclusions}

The De Kee -- Turcotte model has the advantages of the physical significance of its parameters
and its viscous plateau at low shear rates. On the other hand, its exponential shear-thinning
limits its range of applicability: it bounds the magnitude of the stress that it can produce, making
it unusable in high-stress flows. Furthermore, its flow curve exhibits a maximum which splits it
into a stable (stress-increasing) part and an unstable (stress-decreasing) part. The behaviour of
the model in the stable region is akin to the other simple viscoplastic models, such as the
Herschel-Bulkley. An analytical solution was given for planar Poiseuille flow in terms of the
Lambert W function.

Ironically, the model's exponential shear-thinning behaviour with the resulting limitation that it
imposes on the model allows solutions only in cases with mild shear-thinning. Hence, the velocity
profiles in Figs.\ \ref{sfig: velocity profile y0=0 W0} and \ref{sfig: velocity profile y0=0.5 W0}
are reminiscent of mildly shear-thinning power-law and Herschel-Bulkley profiles. However, power-law
and Herschel-Bulkley fluids can undergo much more shear-thinning than a De Kee fluid. In any case
the
viscosity $\eta_1 e^{-t_1 \gd}$ cannot drop below $1/e \approx 36.8\%$ of its zero-shear-rate
value $\eta_1$ if we are to remain in the stable region (Fig.\ \ref{fig: flow curve viscosity}).

One can try to increase the range of applicability by decreasing the time constant $t_1$
(increasing the critical rate of strain $\gd_1$), but this will expand the Newtonian plateau (Fig.\
\ref{fig: flow curve viscosity}) making the fluid more Newtonian (or more Bingham-like in the
viscoplastic case). On the other hand, another possibility for extending the model's range of
applicability would be to incorporate multiple viscous components $\sum_i \eta_i e^{-t_i \gd}$
\cite{DeKee_1980, Kaczmarczyk_2023}. Also, we did not discuss the shear-thickening case, which is
achieved by using negative time constants $t_1 < 0$ (critical rates-of-strain $\gd_1 < 0$). In this
case the stress can grow without bound and the limitation vanishes. For the Poiseuille flow, this
has the implication that in Eq.\ \eqref{eq: dudy} the argument of the $W$ function is now positive
and therefore there are no limitations concerning its magnitude (Fig.\ \ref{fig: Lambert function}).

The non-monotonicity of the De Kee flow curve establishes the existence of unstable solutions
alongside the stable ones, whenever there are solutions at all. The unstable velocity profiles in
planar Poiseuille flow were seen to exhibit inverted and unrealistic features compared to the
stable ones.


\nocite{}
\bibliographystyle{ieeetr}
\bibliography{DeKee}

\end{document}